# The gravitational potential energy of Khufu's pyramid

Charles Hirlimann

Institut de physique et chimie des matériaux de Strasbourg (IPCMS), UMR7504 CNRS-Unistra, 23 rue du Lœss BP 43, F-67034 Strasbourg cedex 2, France.

**Summary**

The calculation of the energy required to raise a constituent block of Khufu's pyramid and the knowledge of the mechanical energy that a worker in charge of the lifting of the blocks can produce per working day allows an estimate of the minimum number of workers necessary for the establishment of a base of the pyramid. It emerges that this number is limited to a maximum of less than 2,000 workers over a working season, i.e. less than 200 workers present each day simultaneously on the site. In addition, the calculation of the flow of blocks leads to the ineluctable conclusion that the construction could only be performed in a massively parallel working mode, undermining the very often put forward hypothesis of the use of ramps to convey the blocks to their final destination.

Khufu's pyramid is in fact a gigantic accumulation of gravitational potential energy. Each block of mass M (kilograms) has a potential energy E (Joules) equal to the product of the gravitational constant on earth g = 9.81 (m / s²) by its altitude h (meters).

One can perfectly calculate this gravitational potential energy of the pyramid, but to do so it is necessary to build a model of the said pyramid.

## Modelling the pyramid

**Pyramid**
H = 140 m, b = 231 m
N = 203 bases

**Block**
Thickness   e = 0.69 m
Width       l = 1.5 x e
Length      L = 2 x e
Area        s = 1.4283 m²

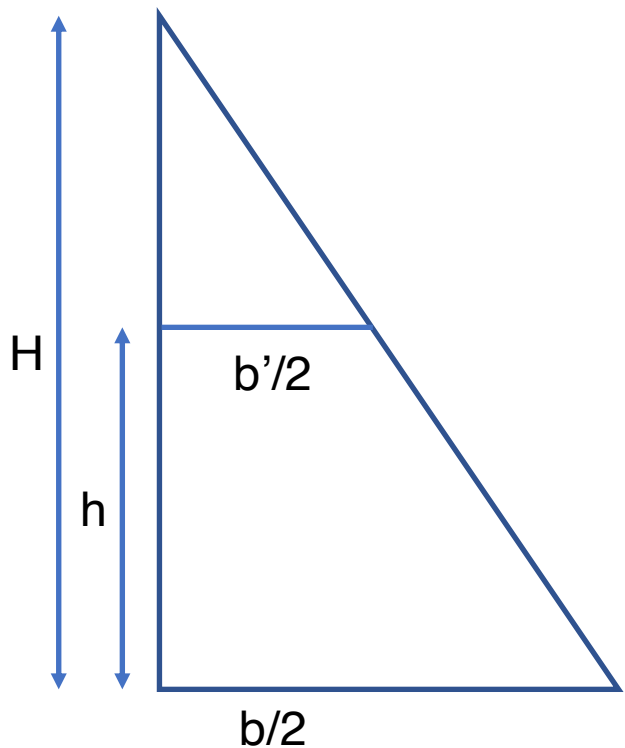

*Figure 1. Half-cross section modeling the pyramid of Khufu. b is the length of the side of the base, b' the length of the side of the course at altitude h. H, is the height of the pyramid. e, the thickness of a block, which has been taken equal to the average thickness of the courses. The blocks have a width of 1.5 times the thickness and a length of 2 times the thickness. s is the floor area of a block.*

Figure 1 shows the choices made to model the pyramid. The height, the length of the side of the base and the average thickness of the courses are taken from values found in the literature. The blocks are modeled with a model block of thickness equal to the average thickness of the courses, 0.69 m, having a width equal to 1.5 times this thickness and a length 2 times this thickness.

*Altitude of course n*

The altitude h of course n, measured as the altitude of the underside of its blocks, is:

$$h = (n-1) * e \qquad (1)$$

and therefore in this reference frame, the altitude of the first course, the one that rests on the ground of the Giza plateau, is h = 0.

*Calculation of the number of blocks per course*

On Figure 1, the two triangles with heights H and h are similar so that:

$$\frac{H-h}{H} = \frac{b'}{b} \tag{2}$$

The value b' of the course side at altitude h is obtained by combining expressions (1) and (2):

$$b' = b\left(1 - \frac{(n-1)*e}{H}\right) \tag{3}$$

The number of blocks in a course is then simply obtained by dividing the surface area of the course by the floor area of one block:

$$N_n = \left(\frac{b'^2}{s}\right) \tag{4}$$

this number must be expressed according to the known quantities and one then obtains:

$$N_n = \frac{b^2}{s}\left(1 - \frac{(n-1)*e}{H}\right)^2 \tag{5}$$

For Khufu's pyramid this leads to the numerical values:

$$\frac{e}{H} = \frac{1}{203} \quad \text{et} \quad N_1 = \frac{b^2}{s} = 37360 \tag{6}$$

Calculating the number of blocks in the course of rank n is easily performed with the following formula:

$$N_n = 37360\left(1 - \frac{(n-1)}{203}\right)^2 \tag{7}$$

In the formula the calculation results are rounded to unity.

## Potential energy calculations

Equipped with formulas (1) for the altitude of the courses and (7) for the number of blocks they contain, we can perform the potential energy calculations by remembering that the gravitational potential energy of a single block at altitude h is writes as:

$$E_p = M * g * h \tag{8}$$

where M is the mass of the block, g the gravitational constant on Earth.

---

The Earth's gravitational constant has the value g = 9.81 m s$^{-2}$. The mass M of the block is the product of its volume by the value of its density $\rho$. The volume is calculated from the numerical values indicated in Figure 1 and is worth V = 0.9855 m$^3$. The density of the limestones used in the pyramid is of the order of 2537 kg / m$^3$ which sets the mass M of the blocks close to 2.5 tonnes.

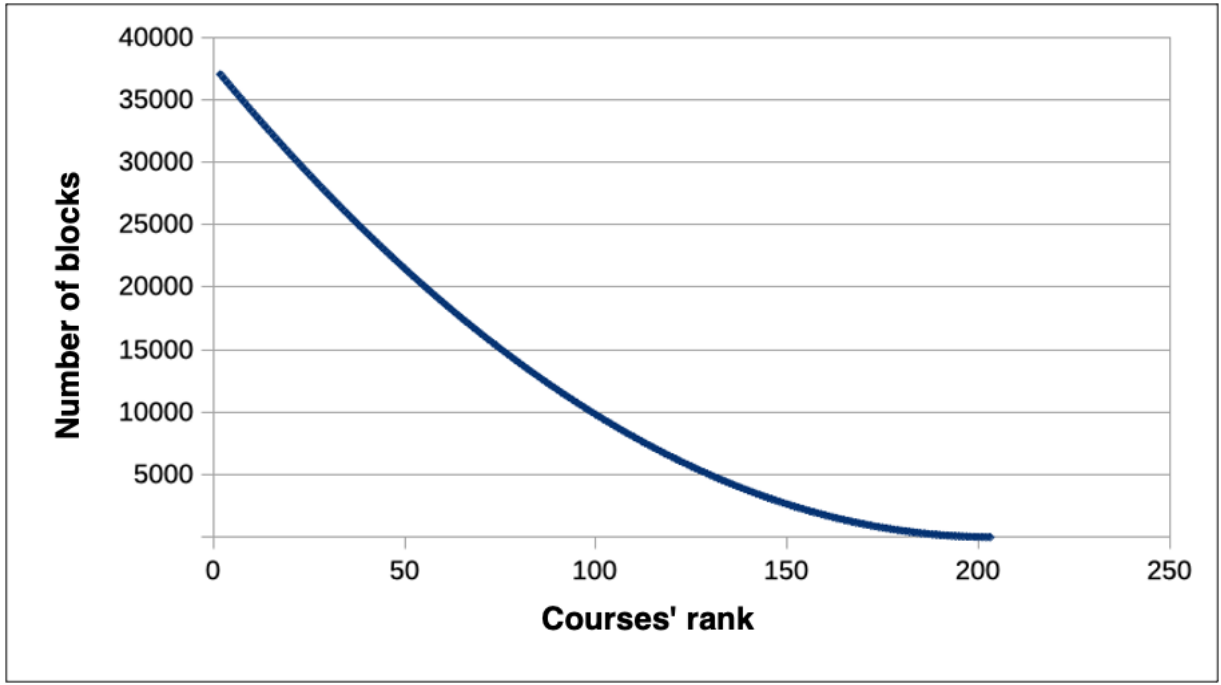

*Figure 2. Variation of the number of blocks in the courses of the model pyramid versus their altitude indicated by their rank. The total number of blocks in this model is 2,546,737.*

Figure 2 shows the variation of the number of blocks in the courses of the model pyramid of Figure 1 as a function of their altitude h, according to formula (7) above. As the number of blocks is proportional to the surface of the courses, this variation follows the one of the square of the side size of each course. The higher one rises on the pyramid, the fewer blocks there are in the courses, therefore the decrease of the number of blocks is faster at the bottom of the pyramid than towards its top. The total number of blocks is around two and a half millions in this fictitious pyramid representative of Khufu's Pyramid. Note that the non-linearity of the curve implies that a third of the blocks of the model pyramid are contained in the first 26 courses and that half is reached in the 42nd course.

*Potential energy of one block*

The potential energy can be calculated using expression (8) above where M is the mass of a block. The altitude h is proportional to the rank of the courses and therefore varies linearly as does the potential energy of a block; the potential energy of a block is proportional to the altitude.

*Potential energy of the courses*

The potential energy contained in one course is simply the product of the potential energy of one block (Expression (8)) by the number of blocks in the course (Expression (7)).

$$E_p = 37360 * M * g * h \left(1 - \frac{(n-1)}{203}\right)^2 \tag{9}$$

Figure 3 (blue curve) shows the variation of the gravitational potential energy of the courses of Khufu's pyramid with the real altitude obtained from the series of measurements of the thickness of the courses[i]. The construction "seasons", with their thickness variations, are visible on the curve as breaks in its continuity. The red curve is the result of a theoretical calculation using the constant courses thickness of the model in Figure 1. In order to obtain a good coincidence between the two curves, it is necessary to increase the values of the red curve by 20%. This disagreement is not surprising, it is due to the inevitable discrepancy between the size of the blocks of the model and the average value of the size of the real blocks which is unknown and also most certainly to the discrepancy between the density used for the calculation of the mass of the blocks and the real density of the limestone constituting the blocks. We can therefore observe in Figure 3 that the combination of a linear increase in the potential energy of a block with a quadratic decreasing number of blocks per course leads to a nice bell curve. This curve goes through a maximum at rank 69. Two strong disturbances of the potential curve are observed at courses 35 and 98. The course 35 is located at the height of the ceiling of the entrance to the Grand Gallery and the course 98 is located approximately 5 m above the relieving chamber of the king's chamber[ii]. This therefore corresponds to the interval of altitudes where the construction of the pyramid competes with the construction of internal granite structures, and there is even room, in this interval for an additional structure compatible with the great void observed through the most recent muon tomographic measurements[iii].

This also corresponds to the construction period when the number of blocks per course has significantly decreased.

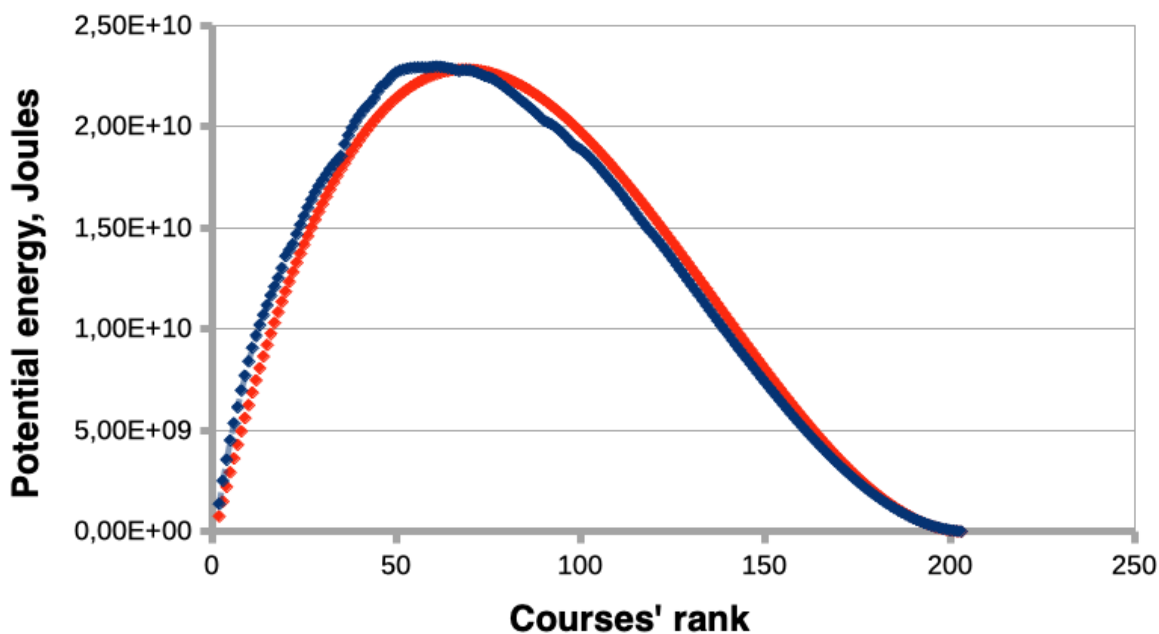

*Figure 3. Gravitational potential energy of the courses: in red the theoretical curve for the model pyramid and in blue the curve obtained using the real altitude of the courses of Khufu's pyramid.*

The total gravitational energy accumulated in the Great Pyramid of Khufuy is approximately 2.5 $10^{12}$ Joules, 2.5 Terajoules which represents only the energy production for 1/2 hour of a 1450 MW nuclear plant!

*The energy source.*

The origin of the potential energy accumulated in the Great Pyramid is by no means mysterious: it is simply made from the animal mechanical energy developed by the workers who did built the monument. Humans can produce a power of a few hundred Watts for a few minutes, but over long periods of time it is reasonable to rely on a lower value. In the following analysis, we will conservatively consider workers working 10 hours a day delivering a power of 50W. Per working day each worker can therefore provide:

$$E_t = 50 * 3600 * 10 = 1{,}8 MJ \tag{10}$$

In the following, this energy will be called a "worker-day".

*Number of worker-days for a block.*

Let us ask ourselves the question of knowing how many workers, working a 10 hour day, (worker-day) are needed to raise a block to the altitude h?

This number of workers is obtained by dividing the potential energy required for a task by the "worker-day" energy:

$$N_t = Mgh \ / \ E_t \tag{11}$$

In our model, this number of workers takes the value:

$$N_t = 0{,}0136 \ h \tag{12}$$

it is therefore possible to calculate the number of workers required per working day to raise a block as a function of its altitude. Calculation shows that the energy produced by two workers in a 10-hour day is enough to lift a 2.5 tonne block to the top of the pyramid.

*Number of worker-days per course and per season*

How many worker-days are needed to raise an entire course? The calculation, for each course, is a simple multiplication of expression (12), evaluated at the altitude h of the concerned course, by the number of blocks in the course (Expression (7)).

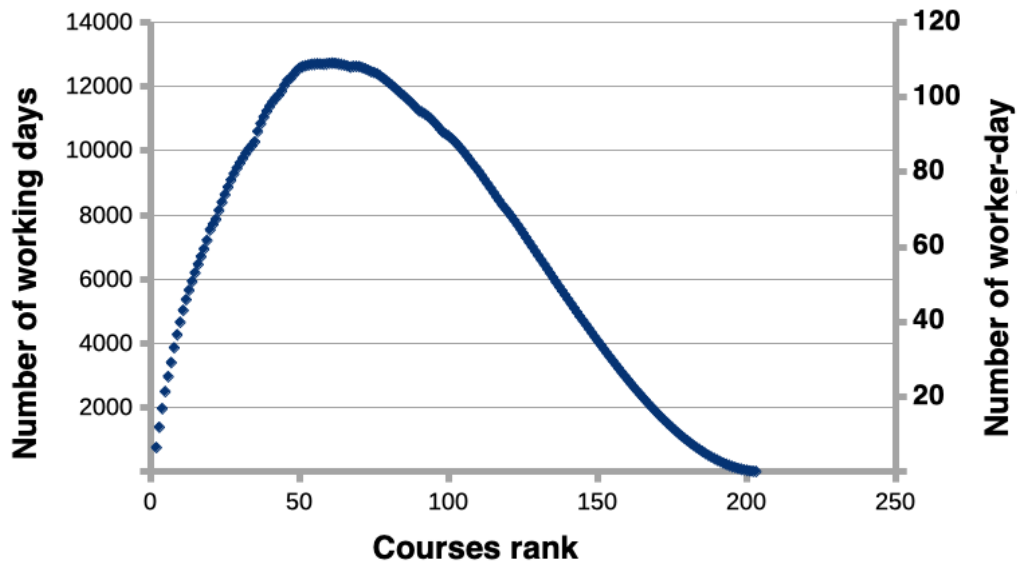

*Figure 4. Variation of the number of worker-days necessary to raise the blocks of the courses. Left scale: total number of 10 hour workdays required. Right scale: number of workers per day in 120-days seasons.*

The result is given in Figure 4. The left scale quantifies the number of minimum working days necessary for the realization of each course of the pyramid by taking as a model a worker producing 50 Watts every second during a 10 hours working day. The maximum energy required is approximately 13,000 working days. This number is important but it should be taken into account that this work is carried out throughout a construction season corresponding to the month of the Nile flood which can be taken to be 120 days long. Over the duration of a season, the number of workers per day therefore becomes on average at most equal to around 110! However, to be more rigorous, the calculation should be done for each "construction season".

*Definition of the seasons*

The division into sections of the thickness of the courses, the "sawtooth" distribution of these thicknesses is a remarkable characteristic of Khufu's pyramid. All the more or less long periods of continuous decrease in the thicknesses of the limestone blocks that constitute them, ends with a strong increase of this thickness. We thus arrive at a fairly widely accepted count of 18 sections.

In this paper we hypothesize that the observable sections in the distribution of the thicknesses of the courses actually reflect the dynamics of the construction of the monument. In other words, each section corresponds to a construction "season" in a year. This value should be compared to the duration of at least 23 years of the reign of Khufu, comparable to the 18 seasons identified; there are thus 5 seasons left for the preparation of the base, the laying of the covering and the construction of the funeral temple so that the pyramid is ready on the death of the pharaoh. This remark leads to the conclusion that each section that can be distinguished in the assembly of the courses must correspond to an annual working season of the peasants summoned during the period of the Nile flood when work in the field is prevented. It therefore seems to us that the denomination of "season" is better suited for the sections of blocks mentioned so far, in that it inserts the static description of the construction into a temporal dynamics.

Figure 5 shows the result of calculating the number of workers whose energy expended on a 10-hour long day for 120 days is needed to build the courses for each season. The values obtained are low, not exceeding 1,800 workers. Of course, these values are not to be taken in absolute value, the energy of many other workers having been necessary for the performance of other tasks in support of the main task. However, these relatively low values show that the erection of the pyramids was not miraculous, but was the coordinated work of a limited number of workers.

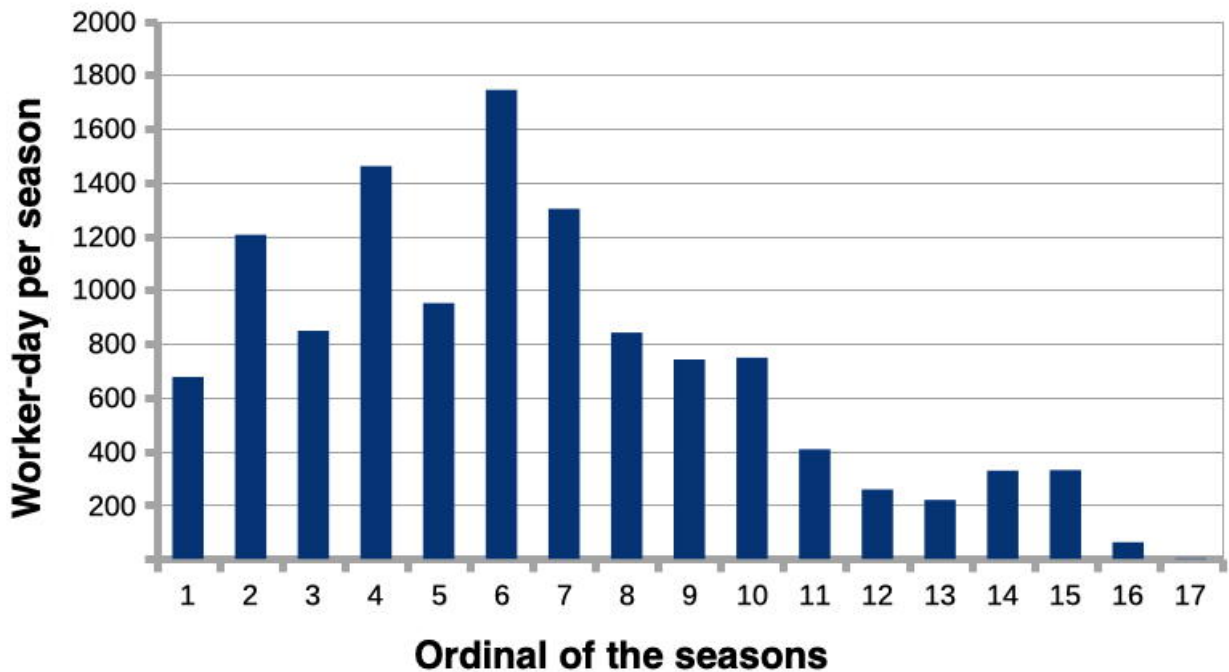

*Figure 5. Total potential energy of the seasons of construction of Khufu's pyramid measured in number of working-days of 10 hours of workers developing a power of 50W.*

### Examining the flows of blocks

Our calculations allow the estimate of the total number of blocks contained in Khufu's pyramid from the expression (7):

$$N_t = \sum_1^{203} 37360 \left(1 - \frac{(n-1)}{203}\right)^2 \tag{13}$$

This gives a value close to 2.5 million blocks. This value is only an approximation which does not take into account the internal structures of the pyramid, nor the rocky core located at its base. It will be corrected as more information becomes available on the internal structure of the pyramid. These 2.5 million blocks had to be put in place in 18 seasons of 120 days in our model, or 2,160 days. Each day, on average, 1,157 blocks had to be put in place permanently. By keeping the 10 hours working day this means that a block had to be placed every 30 seconds or so !!! Such a value is quite impressive and has given rise to many fantasies.

Realism makes it necessary to consider the average flow of blocks for each of the seasons separately. Table 1 gives the result of the calculation of the time interval between the final installation of two consecutive blocks. Up to at least course N° 7 this time interval is less than 30 seconds. It should be qualified by noting that the very first foundations are incomplete in that they include the protuberance of the Giza plateau which has not been flattened. The shortest times correspond to filling more than half of the full building (row 42, end of the third season).

*Table 1: Time interval between the final laying of two successive blocks for each laying season. Until season 7 this time frame is very short and involves a massively parallel working mode.*

| Time s | 7 | 9 | 19 | 14 | 26 | 17 | 26 | 46 | 57 | 61 | 120 | 197 | 240 | 168 | 230 | 1015 | 23358 |
|---|---|---|---|---|---|---|---|---|---|---|---|---|---|---|---|---|---|
| Rank | 1 | 2 | 3 | 4 | 5 | 6 | 7 | 8 | 9 | 10 | 11 | 12 | 13 | 14 | 15 | 16 | 17 |

The time interval substantially increases starting from the eighth course, when there are only about 400,000 blocks left to place.

To achieve such a flow of blocks during the early periods of construction, Egyptian builders had no choice but to perform a massive parallel working mode. A great deal of studies has focused on the methods of construction of the pyramid of Khufu which it is not necessary to detail here. Access ramps of various types have been designed and proposed. Whatever their orientation, parallel or perpendicular to the faces of the pyramid, they can only be in a limited number (essentially 3 or 4) and would not allow the necessary high flows to be achieved, besides their construction and deconstruction would have consumed a large number of worker-days, at a time when this number is already important for the installation of the courses blocks. If we stick to Herodotus[iv] who evokes, without describing them, simple instruments for lifting blocks from one course to the next, one can then imagine the means necessary for achieving high blocks flows. In fact, let us consider, as an example, the altitude of the first two cumulative seasons, rank 34, that we have already encountered and which corresponds to the entrance to the Grand Gallery and therefore to a period of accumulation that does not too much compete with the working related to the construction of the internal structures of the pyramid. At this altitude, 29 m, the side of the course is 162 m. It is assumed here that teams responsible for lifting the blocks from one course to the next are distributed on each side. Along the 162 m length of one side of the course, one can very certainly place ten teams each occupying a space of ten meters which leaves enough room (62 m) to accommodate for the movement of the teams responsible for moving the equipment down, the teams responsible for bringing additional materials to the top, and for quality control teams. Each of the lifting teams then has 40 times longer durations between two block deposits at the top. This value of 40 is only an illustration, the real organization remains unknown and one can think that the arrangements made by the architects of the pyramid might have led to much larger values. This, of course, also involves solving formidable blocks supply management issues. In particular, one must undoubtedly consider that the work of extracting blocks from nearby quarries was spread over the whole year and that a storage of these blocks had been designed all around the foot of the pyramid.

This point of view certainly does not apply to the elevation of the granite elements constituting the internal structure. These elements, very heavy, but few in number may have been raised on a single ramp along the use use of the Great Gallery ramp[v].

It is good at this point to place Occam's razor or the principle of parsimony, which requires limiting as much as possible the number of ad hoc, undocumented hypotheses necessary for the construction of a scenario or a theory. The choice of the two hypotheses, "seasons" and Herodotus' historical note responds all the more to this principle as they are partially documented.

## Conclusion

A simple mathematical analysis, based on a model which is itself simple of Khufu's pyramid, makes it possible, through the means of the gravitational potential energy, to estimate the minimum number of workers necessary, every, for the elevation of the blocks constituting the building. It turns out that at the height of the activity, during the elevation of the first 40 courses, less than 110 workers could provide the energy needed to raise the blocks! The consideration of the flow of blocks necessary to respect the construction deadlines which go down to a few tens of seconds to deposit a block at the height of the activity leads to the inevitable conclusion of a work of raising the blocks practiced in a very strongly parallel way. Only the work of many teams on the faces of the pyramid equipped with simple instruments is compatible with massively parallel processing, as Herodotus reported.

---

[i] HAL: https://hal.archives-ouvertes.fr/hal-02381729
[ii] https://www.cheops-pyramide.ch/khufu-pyramid/stonecourses-pyramid.html
[iii] Discovery of a big void in Khufu's pyramid by observation of cosmic-ray muons, K. Morishima et al, Nature, 552, 386 (2017).
[iv] Mark Lehner, *The Complete Pyramids*, Thames and Hudson, 1997, p. 224
[v] Jean-Pierre Houdin, *Khéops, les secrets de la construction de la grande pyramide*, 2006, éd. du Linteau et Farid Atiya Press.